\documentclass[12pt]{article}
\usepackage{ctex}
\usepackage{amssymb,amsmath,color,mathdots}
\usepackage[colorlinks,linkcolor=blue]{hyperref}
\usepackage{geometry}
\usepackage{tabularx}
\usepackage{bm}
\usepackage{authblk}
\allowdisplaybreaks[4]
\usepackage[square, comma, sort&compress, numbers]{natbib}
\usepackage{enumerate}

\begin{document}
	\title{\bf On the Ding and Helleseth's 9th open problem about optimal ternary cyclic codes}
	\author{\small Peipei Zheng, Dong He and Qunying Liao
		\thanks{Corresponding author.
			
			{~~E-mail. qunyingliao@sicnu.edu.cn (Q. Liao), 18383258130@163.com(P. Zheng).}
			
			{~~Supported by National Natural Science Foundation of China (Grant No. 12071321).}	}
	}
	\affil[] {\small(College of Mathematical Science, Sichuan Normal University, Chengdu Sichuan, 610066, China)}
	\date{}
	\maketitle
	\newtheorem{definition}{Definition}[section]
	\newtheorem{lemma}{Lemma}[section]
	\newtheorem{theorem}{Theorem}[section]
	\newtheorem{proposition}{Proposition}[section]
	\newtheorem{corollary}{Corollary}[section]
	\newtheorem{remark}{Remark}[section]
	\renewcommand\refname{References}	
	\renewcommand{\theequation}{\thesection.\arabic{equation}}
	\newcommand{\upcite}[1]{\textsuperscript{\textsuperscript{\cite{#1}}}}
	\catcode`@=11 \@addtoreset{equation}{section} \catcode`@=12
    {\bf Abstract.}  {\small The cyclic code is a subclass of linear codes and has applications in consumer electronics, data storage systems and communication systems as they have efficient encoding and decoding algorithms. In 2013, Ding, et al. presented nine open problems about optimal ternary cyclic codes. Till now, the 1st, 2nd and 6th problems were completely solved, and the 3rd, 7th, 8th and 9th problems were partially solved. In this manuscript, we focus on the 9th problem. By determining the 
    	root set of some special polynomials over finite fields, we give an incomplete answer for the 9th problem, and then we construct two  classes of optimal ternary cyclic codes with respect to the Sphere Packing Bound 
    	basing on some special polynomials over finite fields.} \\

    {\bf Keywords.} {\small Cyclic code, optimal code, ternary code, Sphere Packing Bound}\\
    
    \section{Introduction}
    Let $p$ be a prime and $m$ be a positive integer. Let $\mathbb{F}_p$ and $\mathbb{F}_{p^m}$ denote the finite fields with $p$ and $p^m$ elements, respectively. A linear code $\mathcal{C}$ with parameters $[n, k, d]$ over the finite field $\mathbb{F}_p$ is a $k$-dimensional subspace of $\mathbb{F}_p^n$ with minimum Hamming distance $d$. $\mathcal{C}$ is cyclic if any cyclic shift of a codeword is also a codeword in $\mathcal{C}$. For the case $\operatorname{gcd}(n, p)=1$, a cyclic code $\mathcal{C}$ can be expressed as $\mathcal{C}=\langle g(x)\rangle$, where $g(x)$ is monic. $g(x)$ is called the generator polynomial of $\mathcal{C}$ and $h(x)=(x^{n}-1) / g(x)$ is called the parity-check polynomial of $\mathcal{C}$. The cyclic code is a class of linear codes with applications in both communication systems and consumer electronics. As a subclass of linear codes, cyclic codes have significant applications in coding theory and communication systems. The recent advances and contributions on cyclic codes can be seen in \cite{A1,A2,A3,A4,A5,A6,A7,A8,A21,A22,A23} and the references therein.
    
    Let $\alpha$ be a generator of $\mathbb{F}_{p^m}^*=\mathbb{F}_{p^m} \backslash\{0\}$ and $m_i(x)$ be the minimal polynomial of $\alpha^i$ over $\mathbb{F}_p$, where $1 \leq i \leq p^m-1$. The cyclic code over $\mathbb{F}_p$ with the generator polynomial $m_u(x) m_v(x)$ is denoted by $\mathcal{C}_{(u, v)}$, where $u$ and  $v$ are from the different $p$-cyclotomic cosets. When $p=3$, the ternary cyclic code with parameters $[3^m-1,3^m-1-2 m, 4]$ is 
    distance-optimal with respect to the Sphere Packing Bound \textsuperscript{\cite{A9}}. For the case $u \neq 1$, several classes of optimal ternary cyclic codes $\mathcal{C}_{(u, v)}$ have been proposed \textsuperscript{\cite{A18,A19,A5}}. For the case $u = 1$, Carlet et al. constructed several optimal ternary cyclic codes basing on perfect nonlinear monomials over $\mathbb{F}_{3^m}$ \textsuperscript{\cite{A10}}. Ding et al. constructed some new classes of optimal ternary cyclic codes by using almost perfect nonlinear monomials (APN) and presented nine open problems by using the monomial $x^v$ over $\mathbb{F}_{3^m}$ \textsuperscript{\cite{A11}}. Till now, the 1st, 2nd and 6th problems were completely solved 
    \textsuperscript{\cite{A12,A13,A14}}. Recently, Ye et al. incompletely  solved the 7th problem and presented a counterexample \textsuperscript{\cite{A24}}. The last two problems for some special values of $h$ were studied \textsuperscript{\cite{A15,A16}}. Furthermore, Zha et al. considered a special case for the 3rd problem and obtained some new classes of optimal ternary cyclic codes \textsuperscript{\cite{A17}}.
    
    In this manuscript, we present two counterexamples for the 9th problem and give three classes of optimal ternary cyclic codes by checking the conditions $Q_1$, $Q_2$ and $Q_3$ in Lemma \ref{l13}. One of them is an incomplete answer for the 9th problem. This manuscript is organized as follows. In Section \ref{h1}, we introduce some necessary preliminaries needed. In Section \ref{h2}, we give a class of optimal ternary cyclic codes with parameters $[3^m-1,3^m-1-2 m, 4]$ by determining the root set of some special polynomials over finite fields.  In Section \ref{h3}, we give two classes of optimal ternary cyclic codes $\mathcal{C}_{(1, e)}$ with parameters $[3^m-1,3^m-1-2 m, 4]$ basing on some polynomials over finite fields. In Section \ref{h4}, we conclude the whole manuscript.
	
	\section{Preliminaries}\label{h1}
	In this section, we first introduce the $p$-cyclotomic coset.
	Let $n=p^m-1$, for any integer $i$ with $0 \leq i \leq n-1$, the $p$-cyclotomic coset modulo $n$ containing $i$ is defined by
	$$
	C_i=\{i p^s ~(\bmod~n) \mid s=0,1, \ldots, \ell_i-1\}
	,$$
	where $\ell_i$ is the minimal positive integer such that $p^{\ell_i} i \equiv i~(\bmod~n)$, and $\ell_i$ is the size of $C_i$, denoted by $|C_i|$.
	
	The following lemmas will be used.
	\begin{lemma}[Theorem 3.46, \cite{A20}]\label{l11}
		Let $k$ be a positive integer and $f$ be an irreducible polynomial of degree $l$ over $\mathbb{F}_p$. Then $f$ can be factorized into $d$ irreducible polynomials in $\mathbb{F}_{p^{k}}[x]$ of the same degree $\frac{l}{d}$, where $d=\operatorname{gcd}(k, l)$.
	\end{lemma}
	\begin{lemma}[Lemma 2.1, \cite{A15}]\label{l12}
		For any integer $i$ with $0 \leq i \leq n-1$, we have $\ell_i \mid m$, where $\ell_i$ is the size of $C_i$.
	\end{lemma}
	\begin{lemma}[Lemma 2.1, \cite{A11}]\label{l14}
		For any integer $e$ with $0 \leq e \leq 3^{m}-2$ and $\operatorname{gcd}(e,3^{m}-1)=2$, we have $|C_e|=m$.
	\end{lemma}
	\begin{lemma}[Lemma 4.1, \cite{A25}]\label{A4}
		For any positive integers $s$ and $n$, let $p$ be a prime with $\operatorname{gcd}(p^s-1,n)=1$. If $t\in\mathbb F_{p^s}^*$, then there exists some $\beta\in\mathbb F_{p^s}^*$ such that $t=\beta^n$.
	\end{lemma}

	It's well-known that a ternary cyclic code with parameters $[3^m-1,3^m-1-2 m, 4]$ is optimal with respect to the Sphere Packing Bound. And for any integer $e$ with $1 \leq e \leq 3^m-1$, Ding and Helleseth gave the following sufficient and necessary condition for the optimal ternary cyclic code $\mathcal{C}_{(1, e)}$ .
	\vspace{3mm}
	\begin{lemma}[Theorem 4.1, \cite{A11}]\label{l13}
		Let $e \notin C_1$ and $|C_e|=m$. Then the ternary cyclic code $\mathcal{C}_{(1, e)}$ has parameters $[3^m-1,3^m-1-2 m, 4]$ if and only if the following conditions are satisfied simultaneously:
		
		$Q_1$. $e$ is even;
		
		$Q_2$. the equation $(x+1)^e+x^e+1=0$ has the unique solution $x=1$ in $\mathbb{F}_{3^m}$;
		
		$Q_3$. the equation $(x+1)^e-x^e-1=0$ has the unique solution $x=0$ in $\mathbb{F}_{3^m}$.
	\end{lemma}
	
	\section{The first class of optimal ternary cyclic codes with minimum distance four}\label{h2}
	In this section, we give a class of optimal ternary cyclic codes $\mathcal{C}_{(1, e)}$ with respect to the Sphere Packing Bound, which is an incomplete answer for the 9th problem in \cite{A11}.

	\vspace{0.8em}\noindent {\bf The 9th Open Problem}\textsuperscript{\cite{A11}}
	Let $e=\frac{3^{m-1}-1}{2}+3^{h}+1$, where $0 \leq h \leq m-1$. What are the conditions on $m$ and $h$ under which the ternary cyclic code $\mathcal{C}_{(1, e)}$ has parameters $[3^m-1,3^m-1-2 m, 4]$?
	 
	 Before give our main results, we first present two counterexamples for the above problem as follows.
	 $\\[8pt]$
	 {\bf Example 3.1}\quad Let $m=5, h=m-1=4$ and $e=\frac{3^{m-1}-1}{2}+3^{h}+1=122$. Basing on Magma program, we can factorize $(x+1)^e+x^e+1$ into the product of the irreducible polynomials over $\mathbb{F}_3$ as follows,
	 \begin{align*}
	 	(x+1)^{122}+x^{122}+1=&(x-1)^2(x^5+x^2+x-1)^2(x^5+x^3+x^2-1)^2 \\
	 	&(x^5-x^3+x^2-x-1)^2(x^5-x^3-x^2-1)^2\\
	 	&(x^5+x^4+x-1)^2(x^5+x^4+x^2-x-1)^2\\
	 	&(x^5+x^4-x^3-x-1)^2(x^5+x^4-x^3+x^2-1)^2\\
	 	&(x^5-x^4-x-1)^2(x^5-x^4-x^2+x-1)^2\\
	 	&(x^5-x^4+x^3+x-1)^2(x^5-x^4-x^3-1)^2,
	 \end{align*}
	 thus $(x+1)^e+x^e+1=0$ has 122 solutions in $\mathbb{F}_{3^{5}}$ by Lemma \ref{l11}. From Lemma \ref{l13} $Q_2$, we know that $\mathcal{C}_{(1, e)}$ is not an optimal ternary cyclic code with respect to the Sphere Packing Bound.
	 $\\[8pt]$
	 {\bf Example 3.2}\quad Let $m=7, h=m-1=6$ and $e=\frac{3^{m-1}-1}{2}+3^{h}+1=1094$. Basing on Magma program, we can factorize $(x+1)^e+x^e+1$ into the product of the irreducible polynomials over $\mathbb{F}_3$ as follows,
	 \begin{align*}
	 	(x+1)^{1094}+x^{1094}+1=&(x-1)^2(x^7-x^2-x-1)^2(x^7+x^3+x^2-1)^2(x^7 + x^3 -x^2 + x-1)^2\\
	 	&(x^7 -x^3 + x^2 -x -1)^2(x^7 + x^4 + x^2 -x -1)^2(x^7 + x^4 + x^3-1)^2\\
	 	&(x^7 + x^4 -x^3-x -1)^2(x^7 -x^4 + x^3 -x^2 -1)^2(x^7 -x^4 -x^3-1)^2\\
	 	&(x^7 + x^5 + x^2 -1)^2(x^7 + x^5 + x^3 -x^2 -1)^2(x^7 + x^5 + x^4 + x^3+ x^2\\
	 	& -1)^2(x^7 + x^5 + x^4 -x^3 + x -1)^2(x^7 +x^5+x^4-x^3 + x^2 -x -1)^2\\
	 	&(x^7 + x^5 -x^4-x^2-1)^2(x^7 + x^5 -x^4 + x^3-1)^2(x^7 + x^5 -x^4 + x^3\\
	 	&-x^2-x-1)^2(x^7 + x^5-x^4-x^3 -x -1)^2(x^7+x^5 -x^4 -x^3-x^2 \\
	 	&+ x -1)^2(x^7-x^5-x^2-1)^2(x^7-x^5+x^3-x^2-x-1)^2(x^7-x^5 \\
	 	&+ x^4 + x^3 + x^2-x-1)^2(x^7-x^5 + x^4-x^3 + x^2 + x-1)^2(x^7-x^5\\
	 	&-x^4-1)^2(x^7-x^5-x^4-x^2-x -1)^2(x^7-x^5-x^4+ x^3-x^2 + x\\
	 	&-1)^2(x^7-x^5-x^4-x^3 + x-1)^2(x^7-x^5-x^4-x^3-x^2 -1)^2(x^7\\
	 	&+ x^6 + x^3 + x^2 + x-1)^2(x^7 + x^6-x^3-x-1)^2(x^7 + x^6 + x^4-x\\
	 	&-1)^2(x^7 + x^6 + x^4 + x^3-x^2-1)^2(x^7 + x^6 + x^4-x^3-1)^2(x^7+ x^6\\
	 	& -x^4 + x^2 + x-1)^2(x^7 + x^6-x^4-x^2-x-1)^2(x^7 + x^6-x^4 + x^3\\
	 	&-x^2 + x-1)^2(x^7 + x^6-x^4-x^3 + x^2-x-1)^2(x^7 + x^6 + x^5-1)^2\\
	 	&(x^7 + x^6 + x^5 + x^3-x-1)^2(x^7 + x^6 + x^5 + x^3 + x^2-1)^2(x^7 + x^6\\
	 	& + x^5 + x^4 + x -1)^2(x^7 + x^6 + x^5+ x^4+ x^3-x^2-x-1)^2(x^7 + x^6\\
	 	&+ x^5-x^4 + x^2-1)^2(x^7 + x^6 + x^5-x^4-x^2+ x-1)^2(x^7 + x^6 + x^5\\
	 	&-x^4+ x^3-x^2 -1)^2(x^7 + x^6 + x^5-x^4-x^3 + x^2 + x-1)^2(x^7 + x^6\\
	 	&+ x^5-x^4-x^3-x^2-x)^2(x^7 + x^6-x^5 + x^3 + x-1)^2(x^7 + x^6-x^5\\
	 	&-x^3-1)^2(x^7 + x^6-x^5 + x^4-1)^2(x^7 + x^6-x^5 + x^4 + x^3-x-1)^2\\
	 	&(x^7 + x^6-x^5 + x^4 + x^3-x^2 + x-1)^2(x^7 + x^6-x^5 + x^4-x^3+ x\\
	 	& -1)^2(x^7 + x^6-x^5+ x^4-x^3-x^2-1)^2(x^7 + x^6-x^5-x^4 + x^3+ x^2\\
	 	& + x-1)^2(x^7 + x^6-x^5-x^4-x^3+ x^2-1)^2(x^7-x^6-x^3-x^2 -x\\
	 	&-1)^2(x^7-x^6+ x^4+ x^3 + x^2-1)^2(x^7-x^6 + x^4+ x^3-x^2 + x -1)^2\\
	 	&(x^7-x^6 + x^4-x^3+ x^2 -x-1)^2(x^7-x^6 + x^4-x^3-x^2-1)^2(x^7\\
	 	&-x^6-x^4+ x^2-x-1)^2(x^7-x^6 + x^5 + x^3-x^2-x-1)^2(x^7-x^6 \\
	 	&+ x^5-x^3-x^2 + x-1)^2(x^7-x^6 + x^5 + x^4-x^2 + x-1)^2(x^7-x^6\\
	 	&+ x^5+ x^4 + x^3-x^2-1)^2(x^7-x^6+ x^5-x^4-1)^2(x^7-x^6 + x^5-x^4\\
	 	& + x^2+ x-1)^2(x^7-x^6 + x^5-x^4+ x^3-x-1)^2(x^7-x^6 + x^5-x^4\\
	 	& + x^3 + x^2-1)^2(x^7-x^6+ x^5-x^4-x^3 + x-1)^2(x^7-x^6 + x^5-x^4\\
	 	&-x^3 + x^2-x-1)^2(x^7-x^6-x^5 + x^3-x-1)^2(x^7-x^6-x^5 + x^3\\
	 	&-x^2+ x -1)^2(x^7-x^6-x^5 + x^4+ x^3-x^2-x-1)^2(x^7-x^6-x^5\\
	 	& + x^4-x^3+ x^2-1)^2(x^7-x^6-x^5-x^4-x-1)^2(x^7-x^6-x^5-x^4\\
	 	&+ x^3+ x^2-x-1)^2,
	 \end{align*}
	 thus $(x+1)^e+x^e+1=0$ has 1094 solutions in $\mathbb{F}_{3^{7}}$ by Lemma \ref{l11}. From Lemma \ref{l13} $Q_2$, we know that $\mathcal{C}_{(1, e)}$ is not optimal with respect to the Sphere Packing Bound.
	 
	 \vspace{0.8em} For convenience, in the following Lemmas \ref{A1}-\ref{A2} and Theorem \ref{A3}, we assume that $h$ is an integer with the prime $m \geq 5$, $0 \leq h \leq m-1$ and
	
	(I) $m\neq5$, $2h\equiv3$ $(\bmod~m)$, i.e., $h=\frac{m+3}{2}$;
	
	\noindent or
	
	(II) $2h\equiv-3$ $(\bmod~m)$, i.e., $h=\frac{m-3}{2}$;

	\noindent or
	
	(III) $m\equiv2$ $(\bmod~3)$ and $3h\equiv1$ $(\bmod~m)$, i.e., $h=\frac{m+1}{3}$.

	\begin{lemma}\label{A1}
		For the prime $m\geqslant5$ and any positive integer $h$, if $e=\frac{3^{m-1}-1}{2}+3^{h}+1$ with $0 \leq h \leq m-1$, then we have $e \notin C_1$ and $|C_e|=m$.
	\end{lemma}
	 {\bf Proof.} It's easy to see that $e \notin C_1$ since $e$ is even. Now from Lemma \ref{l12} we have $|C_e| \mid m$, thus $|C_e|=1$ or $|C_e|=m$ since $m$ is prime. 
	 
	 If $|C_e|=1$, then $3(\frac{3^{m-1}-1}{2}+3^{h}+1) \equiv$ $\frac{3^{m-1}-1}{2}+3^{h}+1(\bmod~3^m-1)$, i.e., $3^m-1 \mid 2(\frac{3^{m-1}-1}{2}+3^{h}+1)$. Note that $m\geqslant5$ and  $2(\frac{3^{m-1}-1}{2}+3^{h}+1)=3^{m-1}+2\cdot3^h+1$, thus $3^{m-1}+2\cdot3^h+1 \leq 3^m-1$, so $3^m-1=3^{m-1}+2\cdot3^h+1$, i.e., $3^{m-1}-3^h=1$ , this is impossible since $3^{m-1}-3^h\neq0$ and $2\mid3^{m-1}-3^h$. Hence, $|C_e|=m.\hfill\Box$
	
	 \vspace{0.8em} For an even integer $e>0$, it can be easily checked that $(x+1)^e+x^e+1=0$ has the unique solution $x=1$ in $\mathbb{F}_{3}$	and $(x+1)^e-x^e-1=0$ has  the unique solutions $x=0$ in $\mathbb{F}_{3}$. To check the conditions $Q_2$ and $Q_3$ in Lemma \ref{l13}, we need to show that there is no solutions in $\mathbb{F}_{3^{m}}\setminus \mathbb{F}_{3}$ for the equation
	 $$
	 (x+1)^e=\pm(x^e+1),
	 $$
	 which means that the equation 
	 \begin{equation}\label{B1}
	 	(x+1)^{6e}=x^{6e}+1-x^{3e}
	 \end{equation}
	 has no solutions in $\mathbb{F}_{3^{m}}\setminus \mathbb{F}_{3}$. The following Lemma \ref{A2} gives the answer.
	 
	\begin{lemma}\label{A2}
	 	For $e=\frac{3^{m-1}-1}{2}+3^{h}+1$, the equation
	 	$$
	 		(x+1)^{6e}=x^{6e}+1-x^{3e}
	 	$$
	 	has no solutions in $\mathbb{F}_{3^{m}}\setminus \mathbb{F}_{3}$.
	\end{lemma}
	 {\bf Proof.} Assume that $\theta \in \mathbb{F}_{3^m} \backslash \mathbb{F}_3$ is a solution for (\ref{B1}). Then we have the following two cases depending on that $\theta$ is a square element or not in $\mathbb{F}_{3^{m}}$.
	 
	 {\bfseries Case 1}~~When $\theta$ is a square element in $\mathbb{F}_{3^{m}}$.
	 It can be verified that $\theta^{6e}=\theta^{2\cdot3^{h+1}+4}$, $\theta^{3e}=\theta^{\frac{3^m-3}{2}+3^{h+1}+3}=\theta^{3^{h+1}+2}$ and
	 $$
	 \begin{aligned}
	  (\theta+1)^{6e} =&(\theta+1)^{4+2\cdot3^{h+1}}=(\theta^{3}+1)(\theta+1)(\theta^{2\cdot3^{h+1}}-\theta^{3^{h+1}}+1)\\
	  =&\theta^{2\cdot3^{h+1}+4}+\theta^{2\cdot3^{h+1}+3}+\theta^{2\cdot3^{h+1}+1}+\theta^{2\cdot3^{h+1}} \\
	  &-\theta^{3^{h+1}+4} -\theta^{3^{h+1}+3}-\theta^{3^{h+1}+1}-\theta^{3^{h+1}} \\
	  &+\theta^{4}+\theta^{3}+\theta+1,
	 \end{aligned}
	 $$
	 thus (\ref{B1}) is equivalent to 
	 $$
	 \begin{aligned}
	 	&\theta^{2\cdot3^{h+1}+3} +\theta^{2\cdot3^{h+1}+1}+\theta^{2\cdot3^{h+1}}-\theta^{3^{h+1}+4}-\theta^{3^{h+1}+3}\\
	 	&+\theta^{3^{h+1}+2}-\theta^{3^{h+1}+1}-\theta^{3^{h+1}}+\theta^{4}+\theta^{3}+\theta=0 ,
	 \end{aligned}
	 $$
	 namely,
    \begin{equation}\label{B2}
	  \theta^{2\cdot3^{h+1}}(\theta^{3}+\theta+1)-\theta^{3^{h+1}}(\theta^{4}+\theta^{3}-\theta^{2}+\theta+1)+\theta^{4}+\theta^{3}+\theta =0.
    \end{equation}
    If $\theta^{3}+\theta+1=0$, then $\theta^{3}+\theta+1=(\theta-1)(\theta^{2}+\theta-1)=0$, we have $\theta=1$ or $\theta^2+\theta-1=0$. Note that $m\geqslant5$ is an odd prime and $\operatorname {gcd}(2,m)=1$, it then follows that $\theta^{2}+\theta-1$ has no solutions over $\mathbb{F}_{3^{m}}$ from Lemma \ref{l11}. So $\theta^3+\theta+1=0$ implies that $\theta=1$, this is contrary to the assumption $\theta\in\mathbb{F}_{3^{m}}\setminus \mathbb{F}_{3}$. Hence, $\theta^{3}+\theta+1\neq0$.
    
    Now set $\theta^{3^{h+1}}=y$. Then (\ref{B2}) is equivalent to 
    \begin{equation}\label{B3}
    y^{2}(\theta^{3}+\theta+1)-y(\theta^{4}+\theta^{3}-\theta^{2}+\theta+1)+\theta^{4}+\theta^{3}+\theta =0,
    \end{equation}
    it's easy to check that $\frac{\theta^{3}+\theta^{2}+1}{\theta^{3}+\theta+1}$ and $\theta$ are both solutions of (\ref{B3}) in $\mathbb{F}_{3^{m}}\setminus \mathbb{F}_{3}$.
    
    If $y=\theta$, then from $\theta\in\mathbb{F}_{3^{m}}\setminus \mathbb{F}_{3}$, 
     $\theta^{3^{h+1}}=\theta$ and $\theta^{3^m}=\theta$, we have $\operatorname{ord} \theta|\operatorname{gcd}(3^{h+1}-1,3^m-1)$ where $\operatorname{ord}\theta$ is the minimal positive integer with $\theta^{ord\theta}=1$. Note that $m\geq5$ is an odd prime and $h=\frac{m+3}{2}(m\neq5)$ or $h=\frac{m-3}{2}$ or $h=\frac{m+1}{3}$, it then follows that $\operatorname{gcd}(h+1,m)=1$, so $\operatorname{gcd}(3^{h+1}-1,3^m-1)=3^{\operatorname{gcd}(h+1,m)}-1=2$. Thus we can get $\operatorname{ord}\theta=1$ or $2$. If $\operatorname{ord}\theta=1$, then $\theta=1$, this is contrary to the assumption $\theta\in\mathbb{F}_{3^{m}}\setminus \mathbb{F}_{3}$. If $\operatorname{ord}\theta=2$, then $\theta^2=1$ and $\theta\neq1$, so $\theta=-1$, this is contrary to the assumption $\theta\in\mathbb{F}_{3^{m}}\setminus \mathbb{F}_{3}$.
     Hence, $y\neq\theta$.
     
     Therefore, we have
    \begin{equation}\label{B4}
    y=\theta^{3^{h+1}}=\frac{\theta^{3}+\theta^{2}+1}{\theta^{3}+\theta+1}=\frac{\theta^{2}-\theta-1}{\theta^{2}+\theta-1}:=\frac{f(\theta)}{g(\theta)},
    \end{equation}
    where $f(\theta)=\theta^{2}-\theta-1$ and $g(\theta)=\theta^{2}+\theta-1$. 
    
    {\bfseries(1.1)} For $m\neq5$, $2h\equiv3~(\bmod~m)$, i.e., $h=\frac{m+3}{2}$. Note that $\theta^{3^m}=\theta$, we obtain $\theta^{3^{2h+2}}=\theta^{3^{m+5}}=\theta^{243}$. Thus by taking the $3^{h+1}$-th power on both sides of (\ref{B4}), we have 
    \begin{equation}\label{B50}
    	\theta^{243}=(\frac{f(\theta)}{g(\theta)})^{3^{h+1}}=\frac{f(\theta)^{2}-f(\theta)g(\theta)-g(\theta)^{2}}{f(\theta)^{2}+f(\theta)g(\theta)-g(\theta)^{2}}:=\frac{F(\theta)}{G(\theta)},
    \end{equation}
    where $F(\theta)=f(\theta)^{2}-f(\theta)g(\theta)-g(\theta)^{2}$ and $G(\theta)=f(\theta)^{2}+f(\theta)g(\theta)-g(\theta)^{2}$, it then follows from (\ref{B50}) that
    $$
    	\theta^{243}G(\theta)-F(\theta)=\theta^{247}-\theta^{246}+\theta^{244}+\theta^{243}+\theta^{4}+\theta^{3}-\theta+1=0.
    $$
    Basing on Magma program, we know that the left-hand side of the above equation can be factorized into the product of the irreducible polynomials over $\mathbb{F}_{3}$ as follows,
    \begin{align}\label{B6}
    \theta^{243}G(\theta)-F(\theta) =&(\theta+1)(\theta^{6}+\theta^{5}+\theta^{4}+\theta^{3}+\theta^{2}+\theta+1)(\theta^{6}-\theta^{5}-\theta^{3}-\theta+1)\notag \\
    &(\theta^{9}-\theta^{8}+\theta^{6}+\theta^{4}+\theta^{3}+\theta^{2}-\theta+1)(\theta^{9}-\theta^{8}+\theta^{7}+\theta^{6}+\theta^{5}\notag\\
    &+\theta^{3}-\theta+1)(\theta^{18}+\theta^{15}+\theta^{13}+\theta^{12}-\theta^{10}+\theta^{9}+\theta^{8}-\theta^{6}+\theta^{5}\notag\\
    &+\theta^{4}-\theta-1)(\theta^{18}+\theta^{16}-\theta^{15}-\theta^{13}+\theta^{10}+\theta^{8}+\theta^{6}+\theta^{5}-\theta^{4}\notag\\
    &+\theta^{3}+\theta-1)(\theta^{18}+\theta^{16}-\theta^{15}+\theta^{14}+\theta^{13}-\theta^{12}+\theta^{10}-\theta^{9}+\theta^{8}\notag\\
    &-\theta^{7}+\theta^{6}-\theta^{5}-\theta^{3}-\theta^{2}-\theta-1)(\theta^{18}+\theta^{17}-\theta^{14}-\theta^{13}\notag\\
    &+\theta^{12}-\theta^{10}-\theta^{9}+\theta^{8}-\theta^{6}-\theta^{5}-\theta^{3}-1)(\theta^{18}+\theta^{17}+\theta^{16}\notag\\
    &+\theta^{15}+\theta^{13}-\theta^{12}+\theta^{11}-\theta^{10}+\theta^{9}-\theta^{8}+\theta^{6}-\theta^{5}-\theta^{4}+\theta^{3}\notag\\
    &-\theta^{2}-1)(\theta^{18}+\theta^{17}+\theta^{16}-\theta^{15}+\theta^{9}+\theta^{8}+\theta^{5}+\theta-1) (\theta^{18}\notag\\
    &-\theta^{17}-\theta^{13}-\theta^{10}-\theta^{9}+\theta^{3}-\theta^{2}-\theta-1)(\theta^{18}-\theta^{17}+\theta^{15}\notag\\
    &+\theta^{12}-\theta^{9}-\theta^{8}+\theta^{7}-\theta^{3}+\theta^{2}+\theta-1)(\theta^{18}-\theta^{17}-\theta^{15}+\theta^{14}\notag\\
    &-\theta^{13}-\theta^{12}-\theta^{10}-\theta^{8}+\theta^{5}+\theta^{3}-\theta^{2}-1)(\theta^{18}-\theta^{17}+\theta^{16}\notag\\
    &-\theta^{12}-\theta^{10}-\theta^{6}-\theta^{5}+\theta^{4}+\theta^{2}+\theta-1)(\theta^{18}-\theta^{17}-\theta^{16}\notag\\
    &-\theta^{14}+\theta^{13}+\theta^{12}+\theta^{8}+\theta^{6}-\theta^{2}+\theta-1)(\theta^{18}-\theta^{17}-\theta^{16}+\theta^{15}\notag\\
    &-\theta^{11}+\theta^{10}+\theta^{9}-\theta^{6}-\theta^{3}+\theta-1).
    \end{align}
   Now from the prime $m\geqslant7$, we know that (\ref{B6}) has no solutions in $\mathbb{F}_{3^{m}}\setminus \mathbb{F}_{3}$ by Lemma \ref{l11}.
    
    {\bfseries(1.2)} For $2h\equiv-3$ $(\bmod~m)$, i.e., $h=\frac{m-3}{2}$. Note that $\theta^{3^m}=\theta$, we have $\theta^{3^{2h+2}}=\theta^{3^{m-1}}=\theta^{\frac{1}{3}}$. Thus by taking the $3^{h+1}$-th power on both sides of (\ref{B4}), we have 
    \begin{equation}\label{B7}
    	\theta^{\frac{1}{3}}=(\frac{f(\theta)}{g(\theta)})^{3^{h+1}}=\frac{f(\theta)^{2}-f(\theta)g(\theta)-g(\theta)^{2}}{f(\theta)^{2}+f(\theta)g(\theta)-g(\theta)^{2}}:=\frac{F(\theta)}{G(\theta)},
    \end{equation}
    where $F(\theta)=f(\theta)^{2}-f(\theta)g(\theta)-g(\theta)^{2}$ and $G(\theta)=f(\theta)^{2}+f(\theta)g(\theta)-g(\theta)^{2}$. By taking the $3$-th power on both sides of (\ref{B7}), we have 
     \begin{equation}\label{B8}
    	\theta=\frac{f(\theta)^{6}-f(\theta)^{3}g(\theta)^{3}-g(\theta)^{6}}{f(\theta)^{6}+f(\theta)^{3}g(\theta)^{3}-g(\theta)^{6}}:=\frac{S(\theta)}{T(\theta)},
    \end{equation}
      where $S(\theta)=f(\theta)^{6}-f(\theta)^{3}g(\theta)^{3}-g(\theta)^{6}$ and $T(\theta)=f(\theta)^{6}+f(\theta)^{3}g(\theta)^{3}-g(\theta)^{6}$, it then follows from (\ref{B8}) that
    $$
    	\theta T(\theta)-S(\theta)=\theta^{13}+\theta^{12}-\theta^{10}+\theta^{9}+\theta^{4}-\theta^{3}+\theta+1=0.
    $$
    Basing on Magma program, we know that the left-hand side of the above equation can be factorized into the product of the irreducible polynomials over $\mathbb{F}_{3}$ as follows,
     \begin{align}\label{B9}
     \theta T(\theta)-S(\theta)=&(\theta+1)\notag\\
     &(\theta^6+\theta^5+\theta^4+\theta^3+\theta^2+\theta+1)\notag\\
     &(\theta^6-\theta^5-\theta^3-\theta+1).
      \end{align}
    Now from the prime $m\geqslant5$, we know that (\ref{B9}) has no solutions in $\mathbb{F}_{3^{m}}\setminus \mathbb{F}_{3}$ by Lemma \ref{l11}.
    
    {\bfseries(1.3)} For $m\equiv2$ $(\bmod~3)$ and $3h\equiv1$ $(\bmod~m)$, i.e., $h=\frac{m+1}{3}$. Note that $\theta^{3^m}=\theta$, we obtain $\theta^{3^{3h+3}}=\theta^{3^{m+4}}=\theta^{81}$. Thus by taking the $3^{2h+2}$-th power on both sides of (\ref{B4}), we have
    \begin{equation}\label{B10}
    	\theta^{81}=(\frac{F(\theta)}{G(\theta)})^{3^{h+1}}=\frac{F(\theta)^{2}-F(\theta)G(\theta)-G(\theta)^{2}}{F(\theta)^{2}+F(\theta)G(\theta)-G(\theta)^{2}}:=\frac{S(\theta)}{T(\theta)},
     \end{equation}
    where $S(\theta)=F(\theta)^{2}-F(\theta)G(\theta)-G(\theta)^{2}$ and $T(\theta)=F(\theta)^{2}+F(\theta)G(\theta)-G(\theta)^{2}$, it then follows from (\ref{B10}) that
    \begin{align}
    		S(\theta)-\theta^{81}T(\theta)=&\theta^{89}-\theta^{88}-\theta^{87}+\theta^{86}+\theta^{85}-\theta^{84}-\theta^{83}+\theta^{82}+\theta^{81}\notag\\
    		&+\theta^{8}+\theta^{7}-\theta^{6}-\theta^{5}+\theta^{4}+\theta^{3}-\theta^{2}-\theta+1=0.\notag
     \end{align}
     Basing on Magma program, we know that the left-hand side of the above equation can be factorized into the product of the irreducible polynomials over $\mathbb{F}_{3}$ as follows,
      \begin{align}\label{B11}
     	S(\theta)-\theta^{81}T(\theta)=&(\theta+1)\notag\\
     	&(\theta^{88}+\theta^{87}+\theta^{86}+\theta^{84}+\theta^{83}+\theta^{82}+\theta^{80}-\theta^{79}+\theta^{78}-\theta^{77}+\theta^{76}\notag\\
     	&-\theta^{75}+\theta^{74}-\theta^{73}+\theta^{72}-\theta^{71}+\theta^{70}-\theta^{69}+\theta^{68}-\theta^{67}+\theta^{66}-\theta^{65}\notag\\
     	&+\theta^{64}-\theta^{63}+\theta^{62}-\theta^{61}+\theta^{60}-\theta^{59}+\theta^{58}-\theta^{57}+\theta^{56}-\theta^{55}+\theta^{54}\notag\\
     	&-\theta^{53}+\theta^{52}-\theta^{51}+\theta^{50}-\theta^{49}+\theta^{48}-\theta^{47}+\theta^{46}-\theta^{45}+\theta^{44}-\theta^{43}\notag\\
     	&+\theta^{42}-\theta^{41}+\theta^{40}-\theta^{39}+\theta^{38}-\theta^{37}+\theta^{36}-\theta^{35}+\theta^{34}-\theta^{33}+\theta^{32}\notag\\
     	&-\theta^{31}+\theta^{30}-\theta^{29}+\theta^{28}-\theta^{27}+\theta^{26}-\theta^{25}+\theta^{24}-\theta^{23}+\theta^{22}-\theta^{21}\notag\\
     	&+\theta^{20}-\theta^{19}+\theta^{18}-\theta^{17}+\theta^{16}-\theta^{15}+\theta^{14}-\theta^{13}+\theta^{12}-\theta^{11}+\theta^{10}\notag\\
     	&-\theta^{9}+\theta^{8}+\theta^{6}+\theta^{5}+\theta^{4}+\theta^{2}+\theta+1).
     \end{align}
    Now from the prime $m\geqslant5$, we know that (\ref{B11}) has no solutions in $\mathbb{F}_{3^{m}}\setminus \mathbb{F}_{3}$ by Lemma \ref{l11}.
    
    {\bfseries Case 2}~~When $\theta$ is not a square element in $\mathbb{F}_{3^{m}}$. It can be verified that  $\theta^{3e}=\theta^{\frac{3^m-3}{2}+3^{h+1}+3}=-\theta^{3^{h+1}+2},$ 
    then (\ref{B1}) is equivalent to 
    $$
    \begin{aligned}
    	&\theta^{2\cdot3^{h+1}+3} +\theta^{2\cdot3^{h+1}+1}+\theta^{2\cdot3^{h+1}}-\theta^{3^{h+1}+4}-\theta^{3^{h+1}+3}\\
    	&-\theta^{3^{h+1}+2}-\theta^{3^{h+1}+1}-\theta^{3^{h+1}}+\theta^{4}+\theta^{3}+\theta=0 ,
    \end{aligned}
    $$
    that is,
    \begin{equation}\label{B12}
    	\theta^{2\cdot3^{h+1}}(\theta^{3}+\theta+1)-\theta^{3^{h+1}}(\theta^{4}+\theta^{3}+\theta^{2}+\theta+1)+\theta^{4}+\theta^{3}+\theta =0.
    \end{equation}
    In the similar proof as that for {\bfseries Case 1}, we can assert that $\theta^{3}+\theta+1\neq0$. 
    
    Now set $\theta^{3^{h+1}}=y$. Then (\ref{B12}) is equivalent to 
    \begin{equation}\label{B13}
    	y^{2}(\theta^{3}+\theta+1)-y(\theta^{4}+\theta^{3}+\theta^{2}+\theta+1)+\theta^{4}+\theta^{3}+\theta =0,
    \end{equation}
    it's easy to check that $\frac{\theta^{4}+1}{\theta^{3}+\theta+1}$ and $\frac{\theta^{3}+\theta^{2}+\theta}{\theta^{3}+\theta+1}$ are both solutions of (\ref{B13}) in $\mathbb{F}_{3^{m}}\setminus \mathbb{F}_{3}$.
    
    {\bfseries(2.1)} When $y=\frac{\theta^{4}+1}{\theta^{3}+\theta+1}$, we have
    \begin{equation}\label{B14}
    	y=\theta^{3^{h+1}}=\frac{\theta^{4}+1}{\theta^{3}+\theta+1}=\frac{\theta^{2}-\theta-1}{\theta-1}:=\frac{f(\theta)}{g(\theta)},
    \end{equation}
    where $f(\theta)=\theta^{2}-\theta-1$ and $g(\theta)=\theta-1$. 
    
   {\bfseries(2.1.1)} For $m\neq5$, $2h\equiv3~(\bmod~m)$, i.e., $h=\frac{m+3}{2}$. Note that $\theta^{3^m}=\theta$, we have $\theta^{3^{2h+2}}=\theta^{3^{m+5}}=\theta^{243}$. Thus by taking the $3^{h+1}$-th power on both sides of (\ref{B14}), we have
    \begin{equation}\label{B15}
    	\theta^{243}=(\frac{f(\theta)}{g(\theta)})^{3^{h+1}}=\frac{f(\theta)^{2}-f(\theta)g(\theta)-g(\theta)^{2}}{f(\theta)g(\theta)-g(\theta)^{2}}:=\frac{F(\theta)}{G(\theta)},
    \end{equation}
    where $F(\theta)=f(\theta)^{2}-f(\theta)g(\theta)-g(\theta)^{2}$ and $G(\theta)=f(\theta)g(\theta)-g(\theta)^{2}$, it then follows from (\ref{B15}) that
    $$
    \theta^{243}G(\theta)-F(\theta)=\theta^{246}-\theta^{244}-\theta^{4}-\theta+1=0.
    $$
     Basing on Magma program, we know that the left-hand side of the above equation can be factorized into the product of the irreducible polynomials over $\mathbb{F}_{3}$ as follows,
    \begin{align}\label{B16}
    	\theta^{243}G(\theta)-F(\theta) =&(\theta^{6}+\theta^{3}-\theta^{2}-\theta+1)(\theta^{6}-\theta^{4}-\theta^{3}+\theta^{2}-\theta-1)(\theta^{9}+\theta^{8}-\theta^{4}\notag\\
    	&-\theta^{2}+\theta+1)(\theta^{9}-\theta^{8}+\theta^{7}-\theta^{6}+\theta^{5}-\theta^{3}-\theta-1)(\theta^{18}+\theta^{12}+\theta^{10}\notag\\
    	&-\theta^{9}-\theta^{8}+\theta^{6}+\theta^{5}-\theta^{4}-\theta^{2}-\theta-1)(\theta^{18}+\theta^{12}+\theta^{10}-\theta^{9}-\theta^{8}\notag\\
    	&+\theta^{7}-\theta^{4}-\theta^{3}+\theta+1)(\theta^{18}+\theta^{15}-\theta^{14}+\theta^{13}-\theta^{12}-\theta^{11}+\theta^{10}\notag\\
    	&-\theta^{9}+\theta^{8}+\theta^{6}-\theta^{5}-\theta^{3}+1)(\theta^{18}-\theta^{15}-\theta^{14}+\theta^{12}+\theta^{9}+\theta^{8}\notag\\
    	&-\theta^{7}+\theta^{6}-\theta^{5}+\theta^{3}-\theta^{2}+\theta-1)(\theta^{18}+\theta^{16}+\theta^{14}+\theta^{13}-\theta^{12}\notag\\
    	&-\theta^{11}-\theta^{10}+\theta^{9}-\theta^{8}+\theta^{7}-\theta^{6}+\theta^{5}+\theta^{4}+\theta^{3}-\theta^{2}-\theta-1)\notag\\
    	&(\theta^{18}+\theta^{16}+\theta^{15}+\theta^{14}+\theta^{12}-\theta^{11}+\theta^{9}-\theta^{8}+\theta^{7}+\theta^{6}+\theta^{5}-\theta^{4}\notag\\
    	&-\theta^{2}+\theta+1)(\theta^{18}+\theta^{17}+\theta^{15}-\theta^{12}+\theta^{10}-\theta^{9}+\theta^{8}+\theta^{7}-\theta^{6}\notag\\
    	&+\theta^{4}-\theta^{3}-\theta^{2}-\theta+1)(\theta^{18}+\theta^{17}-\theta^{15}-\theta^{14}-\theta^{13}+\theta^{12}-\theta^{11}\notag\\
    	&+\theta^{10}+\theta^{9}-\theta^{8}+\theta^{7}-\theta^{5}-\theta^{4}+\theta^{3}-\theta^{2}+\theta+1)(\theta^{18}+\theta^{17}-\theta^{16}\notag\\
    	&+\theta^{15}+\theta^{14}+\theta^{12}+\theta^{9}+\theta^{8}-\theta^{7}-\theta^{6}+\theta^{5}-\theta^{3}-\theta^{2}+\theta+1)(\theta^{18}\notag\\
    	&-\theta^{17}-\theta^{14}-\theta^{13}-\theta^{12}-\theta^{10}+\theta^{6}+\theta^{3}-\theta^{2}-1)(\theta^{18}-\theta^{17}+\theta^{16}\notag\\
    	&+\theta^{13}+\theta^{12}-\theta^{11}+\theta^{9}+\theta^{8}+\theta^{7}+\theta^{6}-\theta^{5}-\theta^{4}-\theta^{3}+\theta^{2}+\theta-1)\notag\\
    	&(\theta^{18}-\theta^{17}+\theta^{16}+\theta^{15}+\theta^{14}-\theta^{13}+\theta^{12}-\theta^{11}-\theta^{10}+\theta^{9}-\theta^{7}\notag\\
    	&-\theta^{4}+\theta^{3}-\theta^{2}-1).
    \end{align}
    Now from the prime $m\geqslant7$, we know that (\ref{B16}) has no solutions in $\mathbb{F}_{3^{m}}\setminus \mathbb{F}_{3}$ by Lemma \ref{l11}.
    
    {\bfseries(2.1.2)} For $2h\equiv-3$ $(\bmod~m)$, i.e., $h=\frac{m-3}{2}$. Note that $\theta^{3^m}=\theta$, we have $\theta^{3^{2h+2}}=\theta^{3^{m-1}}=\theta^{\frac{1}{3}}$. Thus by taking the $3^{h+1}$-th power on both sides of (\ref{B14}), we can get 
    \begin{equation}\label{B17}
    	\theta^{\frac{1}{3}}=(\frac{f(\theta)}{g(\theta)})^{3^{h+1}}=\frac{f(\theta)^{2}-f(\theta)g(\theta)-g(\theta)^{2}}{f(\theta)g(\theta)-g(\theta)^{2}}:=\frac{F(\theta)}{G(\theta)},
    \end{equation}
    where $F(\theta)=f(\theta)^{2}-f(\theta)g(\theta)-g(\theta)^{2}$ and $G(\theta)=f(\theta)g(\theta)-g(\theta)^{2}$. By taking the $3$-th power on both sides of (\ref{B17}) we have
    \begin{equation}\label{B18}
    	\theta=\frac{f(\theta)^{6}-f(\theta)^{3}g(\theta)^{3}-g(\theta)^{6}}{f(\theta)^{3}g(\theta)^{3}-g(\theta)^{6}}:=\frac{S(\theta)}{T(\theta)},
   \end{equation}
    where $S(\theta)=f(\theta)^{6}-f(\theta)^{3}g(\theta)^{3}-g(\theta)^{6}$ and $T(\theta)=f(\theta)^{3}g(\theta)^{3}-g(\theta)^{6}$, it then follows from (\ref{B18}) that
    $$
    	S(\theta)-\theta T(\theta)=\theta^{12}-\theta^{10}+\theta^{4}+\theta^{3}-1=0.
    $$
    Basing on Magma program, we know that the left-hand side of the above equation can be factorized into the product of the irreducible polynomials over $\mathbb{F}_{3}$ as follows,
    \begin{align}\label{B19}
    	S(\theta)-\theta T(\theta)=(\theta^6+\theta^3-\theta^2-\theta+1)(\theta^6-\theta^4-\theta^3+\theta^2-\theta-1).	
    \end{align}
    Now from the prime $m\geqslant5$, we know that (\ref{B19}) has no solutions in $\mathbb{F}_{3^{m}}\setminus \mathbb{F}_{3}$ by Lemma \ref{l11}.
    
    {\bfseries(2.1.3)} For $m\equiv2$ $(\bmod~3)$ and $3h\equiv1$ $(\bmod~m)$, i.e., $h=\frac{m+1}{3}$. Note that $\theta^{3^m}=\theta$, we have $\theta^{3^{3h+3}}=\theta^{3^{m+4}}=\theta^{81}$. Thus by taking the $3^{2h+2}$-th power on both sides of (\ref{B14}), we can get 
    \begin{equation}\label{B20}
    	\theta^{81}=(\frac{F(\theta)}{G(\theta)})^{3^{h+1}}=\frac{F(\theta)^{2}-F(\theta)G(\theta)-G(\theta)^{2}}{F(\theta)G(\theta)-G(\theta)^{2}}:=\frac{S(\theta)}{T(\theta)},
    \end{equation}
    where $S(\theta)=F(\theta)^{2}-F(\theta)G(\theta)-G(\theta)^{2}$ and $T(\theta)=F(\theta)G(\theta)-G(\theta)^{2}$, it then follows from (\ref{B20}) that
    \begin{align}\label{B21}
    	\theta^{81}T(\theta)-S(\theta)=&\theta^{88}-\theta^{87}-\theta^{86}-\theta^{84}+\theta^{83}+\theta^{82}-\theta^{8}\notag\\
    	&+\theta^{7}+\theta^{6}+\theta^{4}-\theta^{3}-\theta^{2}-1=0.
    \end{align}
     Now from the prime $m\geqslant5$, we know that (\ref{B21}) has no solutions in $\mathbb{F}_{3^{m}}\setminus \mathbb{F}_{3}$ by Lemma \ref{l11}.
    
    {\bfseries(2.2)} When $y=\frac{\theta^{3}+\theta^{2}+\theta}{\theta^{3}+\theta+1}$, we have
    \begin{equation}\label{B22}
    	y=\theta^{3^{h+1}}=\frac{\theta^{3}+\theta^{2}+\theta}{\theta^{3}+\theta+1}=\frac{\theta^{2}-\theta}{\theta^{2}+\theta-1}:=\frac{f(\theta)}{g(\theta)},
    \end{equation}
    where $f(\theta)=\theta^{2}-\theta$ and $g(\theta)=\theta^{2}+\theta-1$. 
    
    {\bfseries(2.2.1)} For $m\neq5$, $2h\equiv3~(\bmod~m)$, i.e., $h=\frac{m+3}{2}$. Note that $\theta^{3^m}=\theta$, we can get $\theta^{3^{2h+2}}=\theta^{3^{m+5}}=\theta^{243}$. Thus by taking the $3^{h+1}$-th power on both sides of (\ref{B22}), we have
   \begin{equation}\label{B23}
    	\theta^{243}=(\frac{f(\theta)}{g(\theta)})^{3^{h+1}}=\frac{f(\theta)^{2}-f(\theta)g(\theta)}{f(\theta)^{2}+f(\theta)g(\theta)-g(\theta)^{2}}:=\frac{F(\theta)}{G(\theta)},
    \end{equation}
    where $F(\theta)=f(\theta)^{2}-f(\theta)g(\theta)$ and $G(\theta)=f(\theta)^{2}+f(\theta)g(\theta)-g(\theta)^{2}$, it then follows from (\ref{B23}) that
    $$
    	\theta^{243}G(\theta)-F(\theta)=\theta^{247}-\theta^{246}-\theta^{243}-\theta^{3}+\theta=0.
    $$
    Basing on Magma program, we know that the left-hand side of the above equation can be factorized into the product of the irreducible polynomials over $\mathbb{F}_{3}$ as follows,
    \begin{align}\label{B24}
    		\theta^{243}G(\theta)-F(\theta) =&\theta(\theta^{6}+\theta^{5}-\theta^{4}+\theta^{3}+\theta^{2}-1)(\theta^{6}-\theta^{5}-\theta^{4}+\theta^{3}+1)(\theta^{9}+\theta^{8}\notag\\
    		&+\theta^{6}-\theta^{4}+\theta^{3}-\theta^{2}+\theta-1)(\theta^{9}+\theta^{8}-\theta^{7}-\theta^{5}+\theta+1) \notag\\
    		&(\theta^{18}-\theta^{15}-\theta^{13}+\theta^{12}+\theta^{10}-\theta^{9}+\theta^{8}-\theta^{7}-\theta^{6}+\theta^{5}-\theta^{4}\notag\\
    		&+\theta^{3}+1)(\theta^{18}+\theta^{16}-\theta^{15}-\theta^{12}+\theta^{8}+\theta^{6}+\theta^{5}+\theta^{4}+\theta-1) \notag\\
    		&(\theta^{18}+\theta^{16}-\theta^{15}+\theta^{14}+\theta^{11}-\theta^{9}+\theta^{8}+\theta^{7}-\theta^{6}+\theta^{5}-\theta^{4}\notag\\
    		&-\theta^{3}-\theta^{2}+\theta-1)(\theta^{18}+\theta^{17}-\theta^{15}-\theta^{14}+\theta^{11}-\theta^{10}-\theta^{9}\notag\\
    		&+\theta^{8}+\theta^{6}+1)(\theta^{18}+\theta^{17}+\theta^{16}+\theta^{14}-\theta^{13}-\theta^{12}+\theta^{10}+\theta^{9}\notag\\
    		&-\theta^{8}-\theta^{6}-1)(\theta^{18}+\theta^{17}+\theta^{16}-\theta^{15}-\theta^{14}-\theta^{13}+\theta^{12}-\theta^{11}\notag\\
    		&+\theta^{10}-\theta^{9}+\theta^{8}+\theta^{7}+\theta^{6}-\theta^{5}-\theta^{4}-\theta^{2}-1)(\theta^{18}+\theta^{17}-\theta^{16}\notag\\
    		&-\theta^{14}+\theta^{13}+\theta^{12}+\theta^{11}-\theta^{10}+\theta^{9}-\theta^{7}+\theta^{6}+\theta^{4}+\theta^{3}+\theta^{2}\notag\\
    		&+1)(\theta^{18}+\theta^{17}-\theta^{16}+\theta^{15}-\theta^{14}-\theta^{13}+\theta^{11}
    		-\theta^{10}+\theta^{9}+\theta^{8}\notag\\
    		&-\theta^{7}+\theta^{6}-\theta^{5}-\theta^{4}-\theta^{3}+\theta+1)(\theta^{18}+\theta^{17}-\theta^{16}-\theta^{15}+\theta^{13}\notag\\
    		&-\theta^{12}-\theta^{11}+\theta^{10}+\theta^{9}+\theta^{6}+\theta^{4}+\theta^{3}-\theta^{2}+\theta+1)(\theta^{18}-\theta^{17}\notag\\
    		&+\theta^{16}-\theta^{15}+\theta^{13}-\theta^{12}+\theta^{11}-\theta^{10}-\theta^{9}-\theta^{6}+\theta^{4}+\theta^{3}-1)\notag\\
    		&(\theta^{18}-\theta^{17}-\theta^{16}+\theta^{15}+\theta^{14}+\theta^{13}-\theta^{12}-\theta^{11}-\theta^{10}-\theta^{9}+\theta^{7}\notag\\
    		&-\theta^{6}-\theta^{5}-\theta^{2}+\theta-1)(\theta^{18}-\theta^{17}-\theta^{16}-\theta^{15}+\theta^{14}-\theta^{12}\notag\\
    		&+\theta^{11}+\theta^{10}-\theta^{9}+\theta^{8}-\theta^{6}+\theta^{3}+\theta+1). 
    	\end{align}
    Now from the prime $m\geqslant7$, we know that (\ref{B24}) has no solutions in $\mathbb{F}_{3^{m}}\setminus \mathbb{F}_{3}$ by Lemma \ref{l11}.
    
    {\bfseries(2.2.2)} For $2h\equiv-3$ $(\bmod~m)$, i.e., $h=\frac{m-3}{2}$. Note that $\theta^{3^m}=\theta$, we can get $\theta^{3^{2h+2}}=\theta^{3^{m-1}}=\theta^{\frac{1}{3}}$. Thus by taking the $3^{h+1}$-th power on both sides of (\ref{B22}), we have 
    \begin{equation}\label{B25}
    	\theta^{\frac{1}{3}}=(\frac{f(\theta)}{g(\theta)})^{3^{h+1}}=\frac{f(\theta)^{2}-f(\theta)g(\theta)}{f(\theta)^{2}+f(\theta)g(\theta)-g(\theta)^{2}}:=\frac{F(\theta)}{G(\theta)},
    \end{equation}
    where $F(\theta)=f(\theta)^{2}-f(\theta)g(\theta)-g(\theta)^{2}$ and $G(\theta)=f(\theta)g(\theta)-g(\theta)^{2}$. By taking the $3$-th power on both sides of (\ref{B25}), we have
    \begin{equation}\label{B51}
    	\theta=\frac{f(\theta)^{6}-f(\theta)^{3}g(\theta)^{3}}{f(\theta)^{6}+f(\theta)^{3}g(\theta)^{3}-g(\theta)^{6}}:=\frac{S(\theta)}{T(\theta)},
    \end{equation}
    where $S(\theta)=f(\theta)^{6}-f(\theta)^{3}g(\theta)^{3}$ and $T(\theta)=f(\theta)^{6}+f(\theta)^{3}g(\theta)^{3}-g(\theta)^{6}$, it then follows from (\ref{B51}) that 
    $$
    	\theta T(\theta)-S(\theta)=\theta^{13}-\theta^{10}-\theta^{9}+\theta^{3}-\theta=0.
    $$
    Basing on Magma program, we know that the left-hand side of the above equation can be factorized into the product of the irreducible polynomials over $\mathbb{F}_{3}$ as follows,
    \begin{align}\label{B26}
    	\theta T(\theta)-S(\theta)=&\theta(\theta^6+\theta^5-\theta^4+\theta^3+\theta^2-1)\notag\\
    	&(\theta^6-\theta^5-\theta^4+\theta^3+1).	
    \end{align}
    Now from the prime $m\geqslant5$, we know that (\ref{B26}) has no solutions in $\mathbb{F}_{3^{m}}\setminus \mathbb{F}_{3}$ by Lemma \ref{l11}.
    
    {\bfseries(2.2.3)} For $m\equiv2$ $(\bmod~3)$ and $3h\equiv1$ $(\bmod~m)$, i.e., $h=\frac{m+1}{3}$. Note that $\theta^{3^m}=\theta$, we obtain $\theta^{3^{3h+3}}=\theta^{3^{m+4}}=\theta^{81}$. Thus by taking the $3^{2h+2}$-th power on both sides of (\ref{B22}), we have 
    \begin{equation}\label{B27}
    	\theta^{81}=(\frac{F(\theta)}{G(\theta)})^{3^{h+1}}=\frac{F(\theta)^{2}-F(\theta)G(\theta)}{F(\theta)^{2}+F(\theta)G(\theta)-G(\theta)^{2}}:=\frac{S(\theta)}{T(\theta)},
     \end{equation}
    where $S(\theta)=F(\theta)^{2}-F(\theta)G(\theta)$ and $T(\theta)=F(\theta)^{2}+F(\theta)G(\theta)-G(\theta)^{2}$, it then follows from (\ref{B27}) that
   	\begin{align}
       S(\theta)-\theta^{81}T(\theta)=&\theta^{89}+\theta^{87}+\theta^{86}-\theta^{85}-\theta^{83}-\theta^{82}+\theta^{81}\notag\\
    &-\theta^{7}-\theta^{6}+\theta^{5}+\theta^{3}+\theta^{2}-\theta=0.\notag
    \end{align}
    Basing on Magma program, we know that the left-hand side of the above equation can be factorized into the product of the irreducible polynomials over $\mathbb{F}_{3}$ as follows,
    \begin{align}\label{B28}
    	S(\theta)-\theta^{81}T(\theta)=&\theta(\theta^{88}+\theta^{86}+\theta^{85}-\theta^{84}-\theta^{82}-\theta^{81}+\theta^{80}\notag\\
    	&-\theta^{6}-\theta^{5}+\theta^{4}+\theta^{2}+\theta-1).
    \end{align}
    Now from the prime $m\geqslant5$, we know that (\ref{B28}) has no solutions in $\mathbb{F}_{3^{m}}\setminus \mathbb{F}_{3}$ by Lemma \ref{l11}.$\hfill\Box$
    
    \vspace{0.8em} According to Lemmas \ref{A1}-\ref{A2}, we can get a partial answer for the 9th problem as follows.
    
    \begin{theorem}\label{A3}
    	For $e=\frac{3^{m-1}-1}{2}+3^{h}+1$, the ternary cyclic code $\mathcal{C}_{(1, e)}$ is an optimal ternary cyclic code with parameters
    	$$
        [3^m-1,3^m-1-2 m, 4].
        $$
    \end{theorem}
    
    \section{The second class of optimal ternary cyclic codes with minimum distance four}\label{h3}
    In this section, by studying some special polynomials over finite fields, we give two classes of optimal ternary cyclic codes $\mathcal{C}_{(1, e)}$ with parameters $[3^m-1,3^m-1-2 m, 4]$ for an odd integer $m$. 
    
    For an even integer $e>0$, it can be easily checked that $(x+1)^e+x^e+1=0$ has the unique solution $x=1$ in $\mathbb{F}_{3}$ and $(x+1)^e-x^e-1=0$ has  the unique solution $x=0$ in $\mathbb{F}_{3}$. To check the conditions $Q_2$ and $Q_3$ in Lemma \ref*{l13}, we need to show that there is no solutions in $\mathbb{F}_{3^{m}}\setminus \mathbb{F}_{3}$ of the equation
    $$
    (x+1)^e=\pm(x^e+1),
    $$
    which means that the equation 
    \begin{equation}\label{C2}
    	(x+1)^{2e}-x^{2e}+x^{e}-1=0
    \end{equation}
    has no solutions in $\mathbb{F}_{3^{m}}\setminus \mathbb{F}_{3}$.
      
     \begin{theorem}\label{A7}
     	For any odd integer $m$ and $e=\frac{3^{m}-1}{2}-3$, the ternary cyclic code $\mathcal{C}_{(1, e)}$ has parameters $[3^m-1,3^m-1-2 m, 4]$.
     \end{theorem}
     {\bf Proof.} It's easy to see that $e \notin C_1$ since $e$ is even. Note that $2\mid \operatorname{gcd}(\frac{3^{m}-1}{2}-3,3^{m}-1)$ and $$\operatorname{gcd}(\frac{3^{m}-1}{2}-3,3^{m}-1)\leqslant2\cdot\operatorname{gcd}(\frac{3^{m}-1}{2}-3,\frac{3^{m}-1}{2})=2\cdot\operatorname{gcd}(3,\frac{3^{m}-1}{2})=2,$$ 
     thus $\operatorname{gcd}(\frac{3^{m}-1}{2}-3,3^{m}-1)=2$. By Lemma \ref{l14} we can conclude that $|C_{e}|=m$, thus the condition $Q_1$ in Lemma \ref{l13} is satisfied.
     
     Now we assume that $\theta\in \mathbb{F}_{3^m}\setminus\mathbb{F}_{3}$ is a solution of (\ref{C2}), then we have the following two cases depending on that $\theta$ is a square element or not in $\mathbb{F}_{3^{m}}$.
     
     {\bfseries Case 1}~~When $\theta$ is a square element in $\mathbb{F}_{3^m}$. It can be verified that $\theta^{2e}=\theta^{-6}$, i.e, $\theta^{e}=\theta^{-3}$. Thus (\ref{C2}) is equivalent to
     \begin{equation}\label{B33}
     	(\theta+1)^{-6}-\theta^{-6}+\theta^{-3}-1=0.
     \end{equation}
     From $\theta\in \mathbb{F}_{3^m}\setminus\mathbb{F}_{3}$, by multiplying  $(\theta+1)^{6}\theta^{6}$ on both sides of (\ref{B33}), we can get
     $$
     	\theta^{6}-(\theta+1)^{6}+\theta^{3}(\theta+1)^{6}-(\theta+1)^{6}\theta^{6}=0.
     $$
     Basing on Magma program, we know that the left-hand side of the above equation can be factorized into the product of the irreducible polynomials over $\mathbb{F}_{3}$ as follows,
     \begin{equation}\label{B35}
     	(\theta-1)^{6}(\theta^{2}+1)^{3}=0.
     \end{equation}
      Now from that $m$ is an odd integer and Lemma \ref{l11}, we know  that (\ref{B35}) has no solutions in $\mathbb{F}_{3^{m}}\setminus \mathbb{F}_{3}$.
     
     {\bfseries Case 2}~~When $\theta$ is a not a square element in $\mathbb{F}_{3^m}$. It can be verified that $\theta^{2e}=\theta^{-6}$, $\theta^{e}=-\theta^{-3}$. Thus (\ref*{C2}) is equivalent to
     \begin{equation}\label{B36}
     	(\theta+1)^{-6}-\theta^{-6}-\theta^{-3}-1=0.
     \end{equation}
     From $\theta\in \mathbb{F}_{3^m}\setminus\mathbb{F}_{3}$, by multiplying $(\theta+1)^{6}\theta^{6}$ on both sides of (\ref{B33}), we can get
     $$
     	\theta^{6}-(\theta+1)^{6}-\theta^{3}(\theta+1)^{6}-(\theta+1)^{6}\theta^{6}=0.
     $$
     Basing on Magma program, we know that the left-hand side of the above equation can be factorized into the product of the irreducible polynomials over $\mathbb{F}_{3}$ as follows,
     \begin{equation}\label{B38}
     	(\theta^{2}+\theta-1)^{3}(\theta^{2}-\theta-1)^{3}=0.
     \end{equation}
     Now from that $m$ is an odd integer and Lemma \ref{l11}, we know that (\ref{B38}) has no solutions in $\mathbb{F}_{3^{m}}\setminus \mathbb{F}_{3}$.
     
     By {\bfseries Cases 1}-{\bfseries 2},  the conditions $Q_2$ and $Q_3$ in Lemma \ref{l13} are satisfied.$\hfill\Box$
     
     \begin{lemma}\label{A5}
     	For any integer $m$ with $m\not\equiv0$ $(\bmod~5)$, we have $\operatorname{gcd} (11,3^m-1)=1$.
     \end{lemma}
     {\bf Proof.} Since $m\equiv0$ $(\bmod~5)$, it can be verified that $$3^m-1\equiv3^{5k}-1\equiv(3^5)^{k}-1\equiv(11\times22+1)^k-1\equiv1-1\equiv0~(\bmod~11),$$ 
     where $k$ is integer. Then we obtain that 
   $$\begin{footnotesize}3^m-1 
   	\equiv\left\{\begin{array}{l}
   		3^{5k+1}-1\equiv3(3^5)^{k}-1\equiv3(11\times22+1)^k-1\equiv3-1\equiv2~(\bmod~11), when~m\equiv1~(\bmod~5);\\
   		3^{5k+2}-1\equiv3^2(3^5)^{k}-1\equiv9(11\times22+1)^k-1\equiv9-1\equiv8~(\bmod~11), when~m\equiv2~(\bmod~5);\\
   		3^{5k+3}-1\equiv3^3(3^5)^{k}-1\equiv5(11\times22+1)^k-1\equiv5-1\equiv4~(\bmod~11), when~m\equiv3~(\bmod~5);\\
   		3^{5k+4}-1\equiv3^4(3^5)^{k}-1\equiv4(11\times22+1)^k-1\equiv4-1\equiv3~(\bmod~11), when~m\equiv4~(\bmod~5).
   	\end{array}\right.
   \end{footnotesize}$$
     From the above, we have $\operatorname{gcd}(11,3^m-1)=1$ when $m\not\equiv0$ $(\bmod~5)$. $\hfill\Box$ 
     
     \vspace{0.8em} By Lemma \ref{A4} and Lemma \ref{A5}, we have the following 
     \begin{corollary} \label{C1}
     For any positive integer $m$ with $m\not\equiv0$ $(\bmod~5)$.
     
     (1) If $t\in\mathbb F_{3^m}^*$, then there exists some $\beta\in\mathbb F_{3^m}^*$ such that $t=\beta^{11}$;
     
     (2) If $t\in\mathbb F_{3^m}^*\setminus \left\{-1\right\}$, then there exists some $\theta$ and $\beta\in\mathbb F_{3^m}$ such that $t+1=\theta^{11}$, $t=\beta^{11}$ and $\theta^{11}=\beta^{11}+1$.
    \end{corollary}
     
      \begin{theorem}\label{A8}
     	For any positive integer $e$ with $1\leq e\leq3^m-2$, any odd integer $m\geqslant7$ with $m\not\equiv0$ $(\bmod~9)$ and $m\not\equiv0$ $(\bmod~5)$, the ternary cyclic code $\mathcal{C}_{(1, e)}$ has parameters $[3^m-1,3^m-1-2 m, 4]$ when $11e\equiv2~(\bmod~3^{m}-1)$.
     \end{theorem}
     {\bf Proof.} Since $11e\equiv2~(\bmod~3^{m}-1)$, it can be verified that $e$ is even, $e \notin C_1$ and $\operatorname{gcd}(e, 3^{m}-1)|2$.
     
     (1) Note that $2\mid \operatorname{gcd}(e, 3^{m}-1)$, we have $\operatorname{gcd}(e, 3^{m}-1)=2$, and then $|C_e|=m$ by Lemma \ref{l14}. 
     
     (2) First, we consider the solutions of the equation $(x+1)^{e}+x^{e}+1=0$. For odd integer $m\not\equiv0$ $(\bmod~5)$, it can be verified that $\operatorname{gcd}(11, 3^{m}-1)=1$ by Lemma \ref{A5}. Now for any $x\in\mathbb{F}_{3^{m}}$, there exists $\theta$, $\beta\in\mathbb{F}_{3^{m}}$ such that $x + 1 = \theta^{11}$ and $x = \beta^{11}$ by Corollary \ref{C1}, and so
     \begin{equation}\label{B39}
     	\theta^{11}-\beta^{11}=1.
     \end{equation}
      Thus the equation 
      $$
      	(x+1)^{e}-x^{e}-1=0
      $$
      is equivalent to
      $$
      	\theta^{11e}-\beta^{11e}=1.
      $$
      According to $11e\equiv2~(\bmod~3^{m}−1)$, the above equation can be reduced to 
     $$
      	\theta^{2}-\beta^{2}=1.
     $$
      Set $y=\theta+\beta$, then the above equation leads to $y\in\mathbb{F}_{3^{m}}^{*}$ and $\theta-\beta=\frac{1}{y}$. Thus we have $\theta=-y-\frac{1}{y}$ and $\beta=-y+\frac{1}{y}$. Plugging them into the equation $\theta^{11}-\beta^{11}=1$, we can get
      $$
      	(-y-\frac{1}{y})^{11}-(-y+\frac{1}{y})^{11}=1,
      $$
      which can be simplified as
     $$
     	y^{20}+y^{11}-y^{4}-1=0.
     $$
      Basing on Magma program, we know that the left-hand side of the above equation can be factorized into the product of the irreducible polynomials over $\mathbb{F}_{3}$ as follows,
      \begin{footnotesize}
      \begin{equation}\label{B40}
      (y-1)^{2}(y^{9}+y^8+y^7+y^6+y^5+y^4+y^3+y^2-1)(y^9+y^8+y^7+y^6+y^5+y^4+y^3+y^2-y+1)=0.
      \end{equation}
      \end{footnotesize}
      Now from $m\not\equiv0$ $(\bmod~9)$ and Lemma \ref{l11}, we know that (\ref{B40}) has no solutions in $\mathbb{F}_{3^{m}}\setminus\mathbb{F}_{3}$. This implies that $y=1$ is the unique solution of (\ref{B40}), thus $(x+1)^{e}-x^{e}-1=0$  has the unique solution $x=0$ in $\mathbb{F}_{3^{m}}$.
      
      (3) Next, we consider the solutions of the equation
       $$
      	(x+1)^{e}+x^{e}+1=0.
      $$
      The above equation is equivalent to
      $$
      \theta^{11e}+\beta^{11e}=-1.
      $$
      According to $11e\equiv2~(\bmod~3^{m}−1)$, the above equation can be reduced to 
      $$
      \theta^{2}+\beta^{2}=-1.
      $$
      Set $\theta-\beta=l$ and $\theta\beta=z$, then the above equation leads to
     \begin{equation}\label{B44}
     	l^2-z=-1.
     \end{equation}
      It can be verified that
      $$(\theta^2+\beta^{2})(\theta^{9}-\beta^{9})=\theta^{11}-\beta^{11}+\theta^{2}\beta^{2}((\theta-\beta)^{5}(\theta+\beta)^{2}+\theta^{2}\beta^{2}(-\theta+\beta)^{3}+\theta^{3}\beta^{3}(\theta-\beta)),$$
      which means that
      \begin{equation}\label{B45}
      	-l^9=1+z^2(l^5(l^2+z)-z^2l^3+z^3l).
      \end{equation}
      Now from (\ref{B44})-(\ref{B45}) we can get
      $$
      	l^{11}-l^9+l^7-l^5-l^3-l-1=0.
      $$
      Basing on Magma program, we know that the left-hand side of the above equation can be factorized into the product of the irreducible polynomials over $\mathbb{F}_{3}$ as follows,
      \begin{equation}\label{B46}
      	(l-1)^5(l^2+l-1)(l^4+l^3-l^2-l-1)=0.
      \end{equation}
      Now from that $m$ is an odd integer and Lemma \ref{l11}, we know that (\ref{B46}) has no solutions in $\mathbb{F}_{3^{m}}\setminus\mathbb{F}_{3}$. This implies that $l=1$ is the unique solution  of (\ref{B46}). Thus we have $z=-1$ by (\ref{B44}). It leads to
      \begin{equation}\label{B47}
      	(1+\beta)\beta=-1,
      \end{equation}
      which means that $\beta=1$, and so $x=\beta^{11}=1$. 
      
      From the above and Lemma \ref{l13}, $\mathcal{C}_{(1, e)}$ is an optimal ternary cyclic code with parameters $[3^m-1,3^m-1-2 m, 4]$.
      $\hfill\Box$
     
     \section{Conclusions}\label{h4}
     In this manuscript, we first give two counterexamples for the 9th problem proposed by Ding and Helleseth \textsuperscript{\cite{A11}}. Secondly, basing on properties and polynomials over finite fields, we obtain three sufficient conditions for the ternary cyclic codes $\mathcal{C}_{(1, e)}$ optimal with respect to the Sphere Packing Bound as follows. 
     
     (1) $e=\frac{3^{m-1}-1}{2}+3^{h}+1$, $m\geq5$ is prime with $m\neq5$ and $h=\frac{m+3}{2}$, or $h=\frac{m-3}{2}$, or $m\equiv2$ $(\bmod~3)$ and $h=\frac{m+1}{3}$;
     
     (2) $e=\frac{3^{m}-1}{2}-3$ and $m$ is an odd integer;
     
     (3) $11e\equiv2~(\bmod~3^{m}-1)$, $m$ is an odd positive integer with
      $m\geqslant7$, $m\not\equiv0$ $(\bmod~9)$ and $m\not\equiv0$ $(\bmod~5)$, $e$ is a positive integer with $1\leq e\leq3^m-2$.
    
     It's easy to see that (1) is just  an incomplete answer for the 9th problem proposed by Ding and Helleseth \textsuperscript{\cite{A11}}.

\end{document}